\documentclass[a4paper]{article}

\usepackage[dvips]{graphicx}
\usepackage{cite}
\usepackage{amsfonts}
\usepackage[cmex10]{amsmath}
\usepackage[caption=false,font=footnotesize]{subfig}

\advance \oddsidemargin by -\textwidth
\divide  \oddsidemargin by 2
\advance \oddsidemargin by -1in
\evensidemargin = \oddsidemargin

\newlength{\figurewidth}
\def\max{\mathrm{max}}
\def\SNR{\mathsf{SNR}}
\def\pSNR{\mathsf{pSNR}}

\begin{document}
\title{Capacity and Modulations with Peak Power Constraint}

\author{Shiro~Ikeda
  \thanks{S. Ikeda is with the Institute of Statistical Mathematics,
    Tokyo, 190-8562, Japan, e-mail: (shiro@ism.ac.jp).},
  Kazunori~Hayashi,
  and~Toshiyuki~Tanaka
  \thanks{K. Hayashi and T. Tanaka are with the Department of Systems
    Science, Kyoto University, Kyoto, 606-8501, Japan, e-mails:
    (\{kazunori, tt\}@i.kyoto-u.ac.jp)}%
}

\maketitle

\begin{abstract}
  A practical communication channel often suffers from constraints on
  input other than the average power, such as the peak power
  constraint. In order to compare achievable rates with different
  constellations as well as the channel capacity under such
  constraints, it is crucial to take these constraints into
  consideration properly. In this paper, we propose a direct approach
  to compare the achievable rates of practical input constellations
  and the capacity under such constraints. As an example, we study the
  discrete-time complex-valued additive white Gaussian noise (AWGN)
  channel and compare the capacity under the peak power constraint
  with the achievable rates of phase shift keying (PSK) and quadrature
  amplitude modulation (QAM) input constellations.
\end{abstract}

\begin{keywords}
  peak power constraint, capacity, AWGN channel, PSK, QAM, APSK.
\end{keywords}

\section{Introduction}

The channel capacity is defined as the supremum of the mutual
information between input and output~\cite{Shannon1948}, where the
supremum is generally taken under some constraint on the input. For a
band-limited channel, a well-known result is the Shannon-Hartley
theorem, which dictates the capacity $W\log(1+\SNR)$ ($\SNR$:
signal-to-noise ratio) for a complex-valued additive white Gaussian
noise (AWGN) channel with a bandwidth of $W$ under the average power
constraint on the input. We do not have to add anything about
theoretical and conceptual importance of this celebrated formula,
found in almost all textbooks on communication theory. One has to be
a bit careful, however, about its practical significance because a
real-world communication system suffers from limitations other
than the average power constraint. For example, from an engineering
viewpoint, the peak power constraint is important, because the power
amplifier of a communication system physically has an absolute peak
power (amplitude) limitation. It should also be noted that power
efficiency of the amplifier largely depends on the peak value of the
continuous-time input signal~\cite{Raab_etal2002ieeemtt}.

Although the importance of the peak power constraint has been
realized, it has mostly been considered only indirectly via the
peak-to-average power ratio (PAR)~\cite{TseViswanath2005cup}. Indeed,
typical conventional arguments define the capacity under the average
power constraint, and discuss the peak power (or the power efficiency)
only via PAR. The most crucial drawback with this indirect approach is
that, under the peak power constraint, the quantity $W\log(1+\SNR)$ is
no longer the capacity and is therefore no more an appropriate
performance measure.

In this paper, we propose a direct approach to study the achievable
rates and the capacity under proper constraints on input. The
practical significance with this direct approach is that it allows us
to evaluate quantitatively how close the achievable rates to the
theoretical limit posed by the capacity under a practical constraint
on input. One can also expect that the proposed approach will provide
a fresh look at the problem of comparing performance between different
input modulation schemes. A major weakness with our approach would be
that one can no longer expect a simple closed-form expression for the
capacity, such as $W\log(1+\SNR)$, so that evaluation of the capacity
itself might be elaborative and computationally intensive. We
nevertheless believe, despite this weakness, the significance of our
approach in view of better understanding of the room for improvement
in achievable rates toward the theoretical limit under practical
constraints. In order to demonstrate the significance of our approach,
we study, as an example, the capacity of a complex-valued AWGN channel
under peak power constraint on discrete-time signal (as in \cite[Sec.\
12.5]{Goldsmith2005cup}) and compare it with the achievable rates with
practical input constellations such as phase shift keying (PSK) and
quadrature amplitude modulation (QAM). We then show how the achievable
rates with PSKs and QAMs behave and reveal that they are surprisingly
close to the capacity. Our results imply that a well-designed
practical discrete adaptive
modulation~\cite{ChungGoldsmith2001ieeecom} can achieve a rate which
is very close to the capacity.

\section{Modulations and Constraints}

\subsection{AWGN Channel and Peak Power Constraints}
\label{subsec:ppc}

We consider a discrete-time complex-valued AWGN channel, which is
assumed memoryless and having an isotropic independent Gaussian noise,
\begin{align}
  \label{eq:awgnc}
  \begin{pmatrix}
    Y_I\\Y_Q
  \end{pmatrix}
  =
  \begin{pmatrix}
    X_I\\X_Q
  \end{pmatrix}
  +
  \frac{\sigma}{\sqrt{2}}
  \begin{pmatrix}
    N_I\\N_Q
  \end{pmatrix},\hspace{1em}
  N_I,N_Q\sim \mathcal{N}(0,1),
\end{align}
where $X_I$ and $X_Q$ denote in($I$)- and quadrature($Q$)-phase input
components, respectively.

\begin{figure}[h]
  \centering
  \includegraphics[width=0.8\figurewidth]{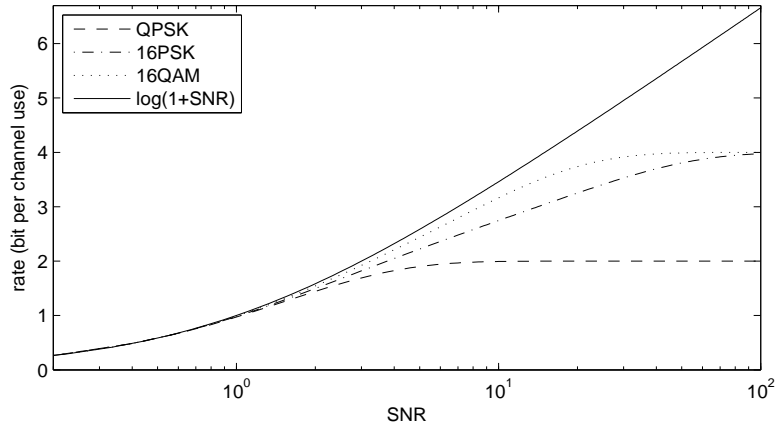}
  \caption{Achievable rates with QPSK, 16PSK, and 16QAM compared with
    capacity under average input power constraint.}
  \label{fig:Modulations_SN}
\end{figure}

For digital communications systems, choosing an appropriate discrete
input-signal constellation is a key issue. Figure
\ref{fig:Modulations_SN} shows the achievable rates with QPSK, 16PSK,
and 16QAM input constellations for the AWGN channel and Shannon's
capacity $\log(1+\SNR)$, which is the capacity for the AWGN channel
under the average input power constraint. One observes from
Fig.~\ref{fig:Modulations_SN} that the achievable rate with 16QAM
input constellation is almost always closer to the capacity than those
with other PSK input constellations. The results shown here imply
that, if we assume a system with adaptive modulation which can use
QPSK, 16PSK, and 16QAM, then such a system should always choose 16QAM,
if we do not think about complexity in implementation.

The above comparison is well-known, but not appropriate in practice
from two reasons. Firstly, for the single-carrier transmitter, the
average power consumption of the amplifier is dominated by
back-off~\cite{Raab_etal2002ieeemtt}, and it is therefore more
reasonable to compare different signal constellations by aligning
their amplitudes in terms not of their average power but of their peak
power. Indeed, for the same average power, the peak power of 16QAM
constellation is 9/5 times larger than that of PSKs, which results in
an unfair comparison under the peak power constraint. Secondly, the
capacity $\log(1+\SNR)$ is achieved only by a Gaussian input
distribution whose support is unbounded. This cannot be realized under
peak power constraint.

\begin{figure}[h]
  \centering
  \subfloat[Box constraint.]{
    \includegraphics[width=0.35\figurewidth]{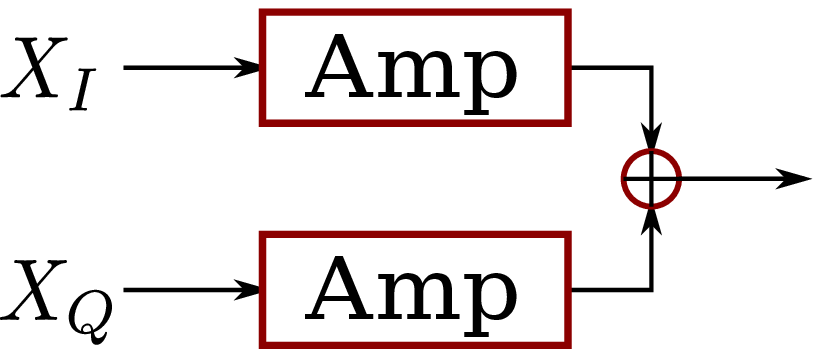}
    \label{fig:Box}
  }
  \subfloat[Circular constraint.]{
    \includegraphics[width=0.35\figurewidth]{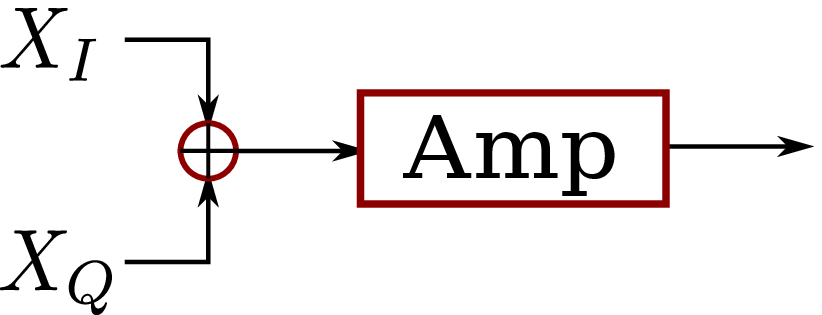}
    \label{fig:Circle}
  }
  \caption{Two types of peak power constraints.}
  \label{fig:Box and Circle}
\end{figure}
There are two different forms of peak power constraint which are
considered natural for wireless digital communications systems,
according to different implementations of the transmitter front-end.
Figure \ref{fig:Box} shows the case in which each component has an
amplifier separately. Assuming that the peak power of each amplifier
is bounded equally, a natural form of peak power constraint is the
component-wise constraint $X_I^2, X_Q^2\le E_{\max}/2$. We call this
the {\em box constraint}. Another implementation is shown in
Fig.~\ref{fig:Circle}, where the sum of the components is amplified at
once. The peak power constraint in this case is formulated as
${X_I^2+X_Q^2}\le E_{\max}$, which we call the {\em circular
  constraint}.

Let $\pSNR=E_{\max}/\sigma^2$ denote the ratio of peak input power to
the noise variance, which we call peak SNR. Under the peak power
constraint, the capacity of the AWGN channel~\eqref{eq:awgnc} must be
smaller than $\log(1+\pSNR)$ for two reasons: 1) $\pSNR\ge\SNR$ holds
and 2) the input distribution cannot be Gaussian. It is known that the
capacity-achieving distribution (CAD) for the AWGN channel under peak
power constraint often becomes discrete. This phenomenon was first
shown for a scalar AWGN channel~\cite{Smith1971informcontrol}, and has
been extended for many channels with different
constraints~\cite{ShamaiBarDavid1995ieeeit,Gursoy_etal2005ieeewc,Chan_etal2005ieeeit,IkedaManton2009nc}.
Using these results, we numerically evaluate the capacity under each
of the box and circular constraints in the following subsections.

\subsection{Box Constraint}

Under the box constraint, the $I$- and $Q$-components of the
channel~\eqref{eq:awgnc} suffer from independent Gaussian channel
noises as well as independent peak power constraints. Accordingly, the
channel~\eqref{eq:awgnc} is decomposed into two independent
real-valued AWGN channels under the respective peak power constraints
$X_I^2\le E_{\max}/2$ and $X_Q^2\le E_{\max}/2$. The capacity of the
complex-valued AWGN channel is thus attained by the direct product of
CADs of the two real-valued AWGN channels under peak power constraint.

\begin{figure}[htbp]
  \centering
  \subfloat[Locations of mass points of CADs for a real-valued scalar
  AWGN channel.]{
    \includegraphics[width=0.80\figurewidth]{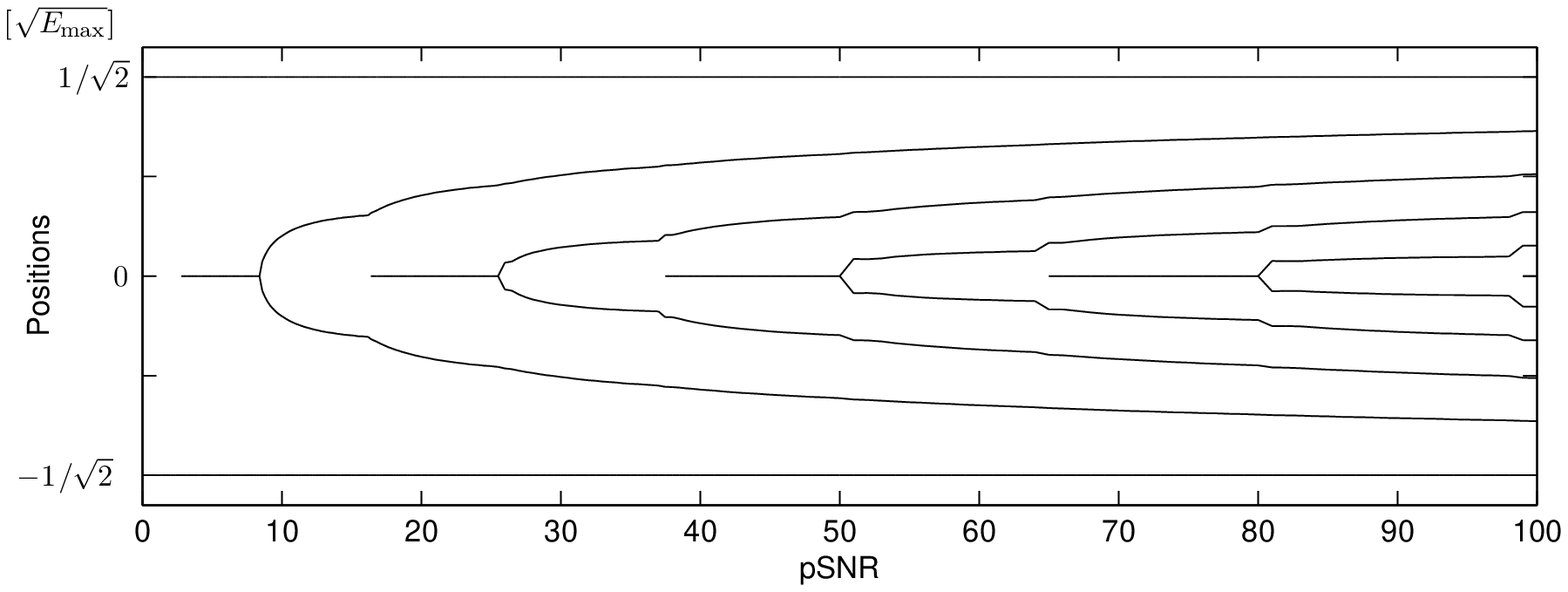}
    \label{fig:box-CAD-a}
  }\\
  \subfloat[CAD when $\pSNR=1$.]{
    \includegraphics[width=0.45\figurewidth]{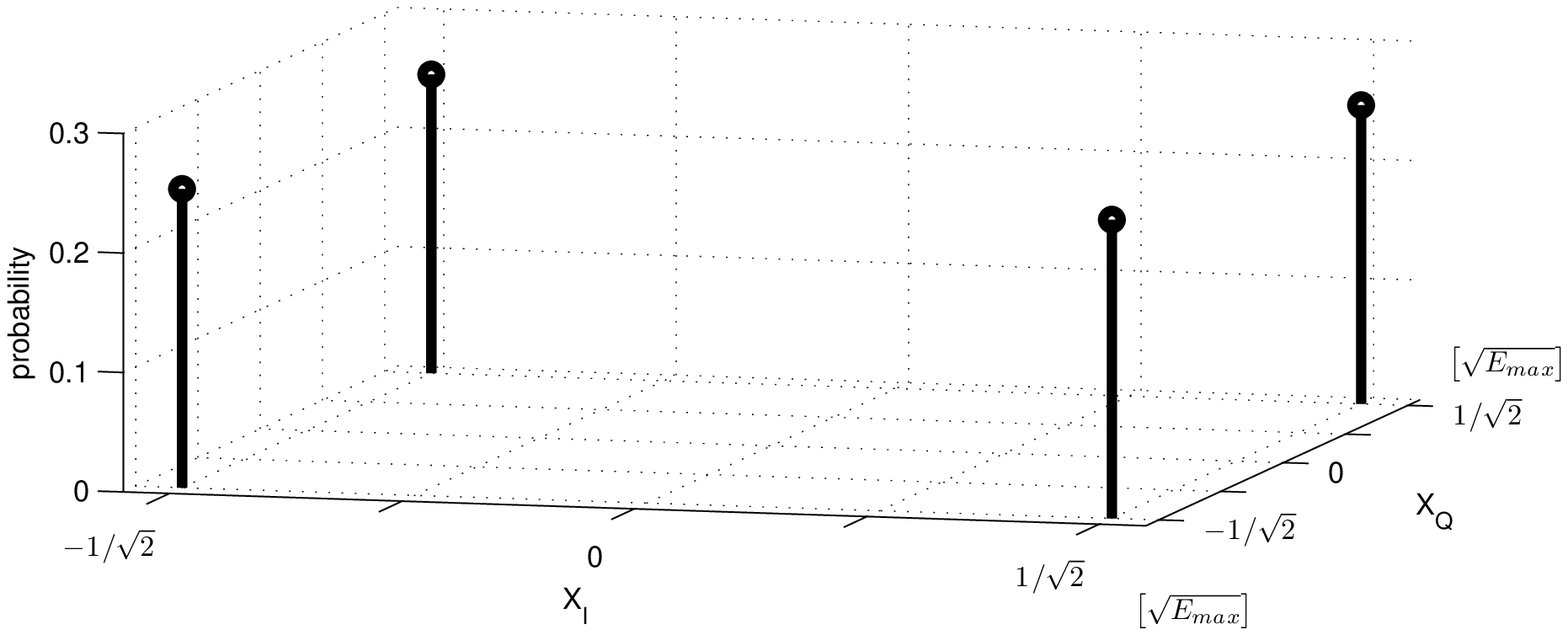}
    \label{fig:box-CAD-b}
  }
  \subfloat[CAD when $\pSNR=16$.]{
    \includegraphics[width=0.45\figurewidth]{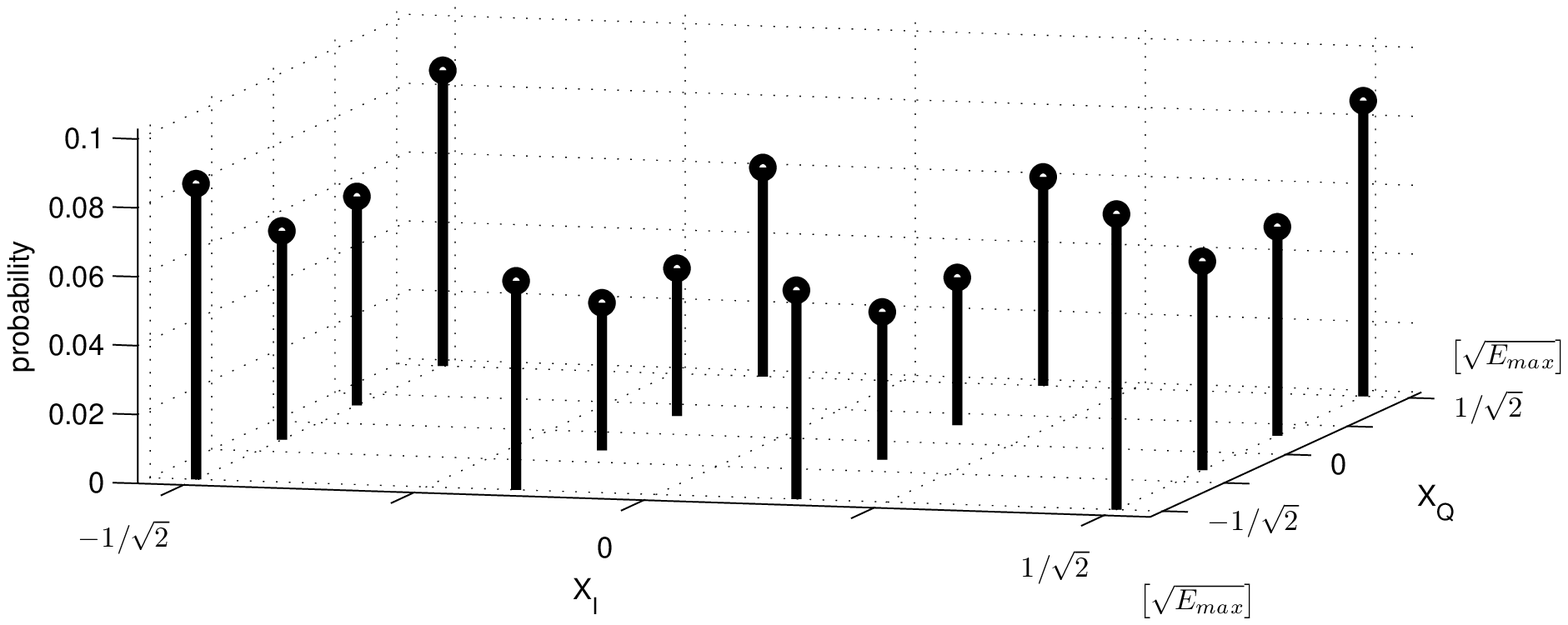}
    \label{fig:box-CAD-c}
  }
  \caption{CADs of AWGN channel under box constraint.}
  \label{fig:box-CAD}
\end{figure}

The capacity of a real-valued scalar AWGN channel under the peak power
constraint has been studied by Smith~\cite{Smith1971informcontrol}. He
has proved that the capacity is achieved by a discrete input
distribution with a finite number of probability mass points. No
analytical solution is known for the capacity itself, nor the CAD. We
evaluated them numerically via the method described
in~\cite{Smith1971informcontrol} with Gauss-Hermite
integration. Figure~\ref{fig:box-CAD-a} shows the positions of the
probability mass points of the CAD versus $\pSNR$. The points are
symmetrically positioned around 0 and two points are always located at
the boundaries $\pm \sqrt{E_{\max}/2}$. The number of the probability
mass points of the CAD is 2 for low enough peak SNRs, while it
increases as $\pSNR$ becomes larger. It is in contrast with the case
under average power constraint, where the CAD is Gaussian and remains
essentially the same irrespective of noise level.

Figures \ref{fig:box-CAD-b} and \ref{fig:box-CAD-c} show CADs for the
complex-valued AWGN channel under the box constraint with different
$\pSNR$ values. They show that QPSK is the CAD for small enough
$\pSNR$ and that 16QAM is similar to the CAD around the $\pSNR$ level
used in Fig.~\ref{fig:box-CAD-c}. The number of the probability mass
points is $m^2$, where $m\ge 2,\;m\in \mathbb{N}$, and probability
masses on these points are generally not equal for $m>2$.

\begin{figure}[htb]
  \centering
  \includegraphics[width=0.8\figurewidth]{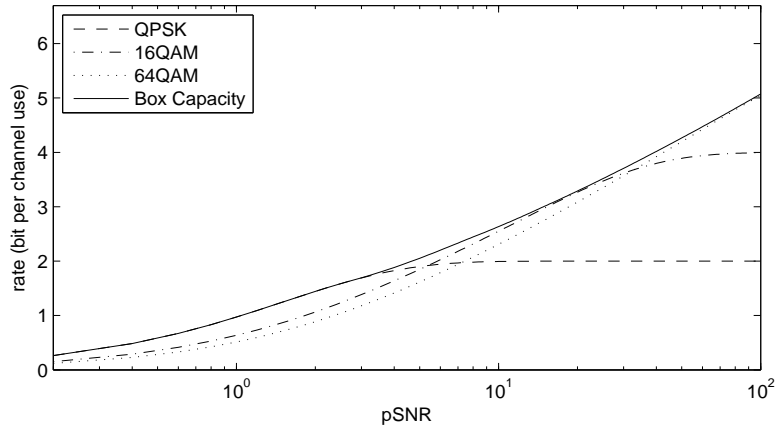}
  \caption{Achievable rates with QPSK, 16QAM, and 64QAM compared
    with capacity under box constraint.}
  \label{fig:QAMs_maxS2N}
\end{figure}
Figure \ref{fig:QAMs_maxS2N} shows the achievable rates with $n$QAMs
and the capacity under the box constraint. One can observe that each
of the achievable rates comes very close to the capacity around
intermediate $\pSNR$ values, and that the $\pSNR$ range in which the
achievable rate with $n$QAM comes close to the capacity shifts
rightwards as $n$ increases. This observation is ascribed to the fact
that the $n$QAMs are similar in their shapes to the CADs under the box
constraint in the respective $\pSNR$ ranges.

The above results also indicate that appropriate switching between
$n$QAMs with different $n$ will achieve rates that are close to the
capacity under the box constraint. For example, among QPSK, 16QAM, and
64QAM, QPSK is the best for $\pSNR<5.4$, 16QAM is the best for $\pSNR$
values between $5.4$ and $33$, and 64QAM is the best for larger
$\pSNR$. From Fig.~\ref{fig:QAMs_maxS2N}, the degradation of the
achievable rate with the above discrete adaptive modulation from the
capacity in terms of $\pSNR$ for the rates 1, 2, and 3 are 0.0, 1.0,
and 0.012 dB, respectively.

\subsection{Circular Constraint}

The capacity and the joint distribution of $X_I$ and $X_Q$ which
achieves the capacity under the circular constraint have been studied
in~\cite{ShamaiBarDavid1995ieeeit}. The result is best described with
the polar coordinate. Reparameterizing $X_I$ and $X_Q$ with the radius
$r$ and the phase $\phi$, the CAD is uniform for $\phi$ and discrete
with a finite number of probability mass points for $r$. Consequently,
the CAD consists of concentric circles centered at the origin. The
number of the circles and their radii, as well as their probability
weights vary with $\pSNR$. The analytical solution is not available,
and we computed the capacity and the CAD numerically via the method
described in~\cite{ShamaiBarDavid1995ieeeit} with Gauss-Laguerre
integration.

\begin{figure}[htbp]
  \centering
  \subfloat[Locations of mass points of CADs in radial coordinate under
  circular constraint.]{
    \includegraphics[width=0.8\figurewidth]{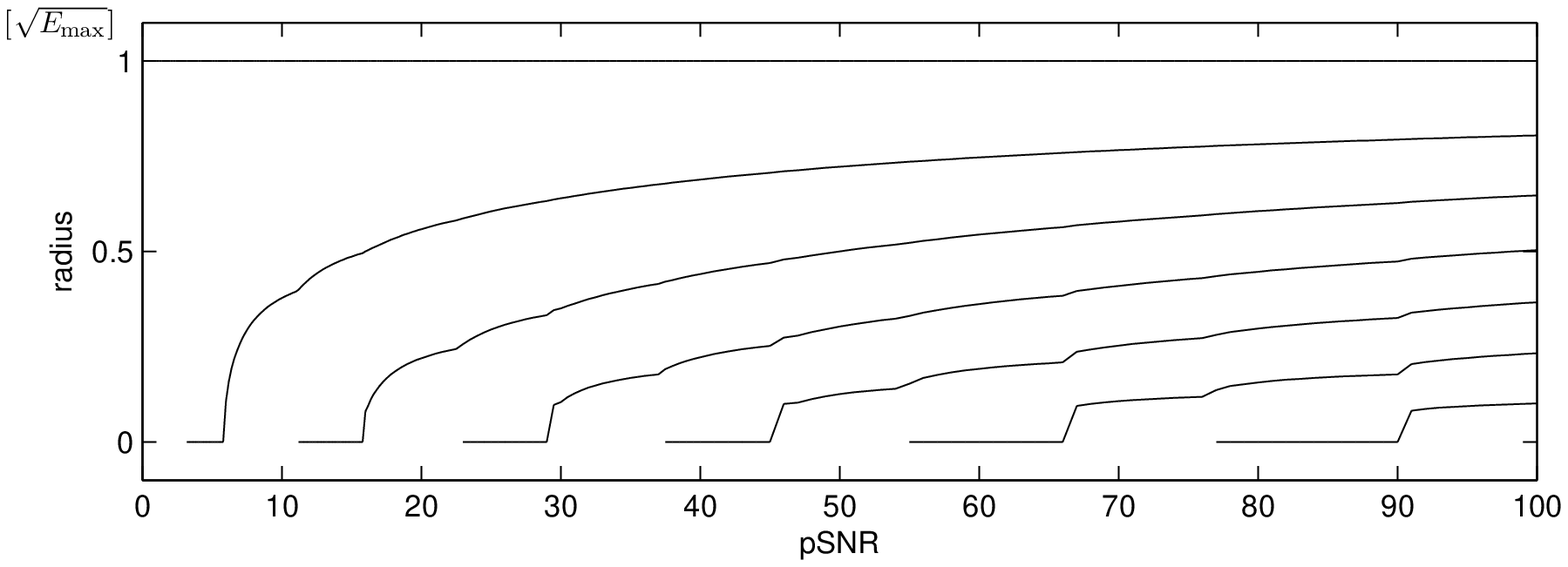}
    \label{fig:circle-CAD-a}
  }
  \\
  \subfloat[CAD when $\pSNR=1$.]{
    \includegraphics[width=0.45\figurewidth]{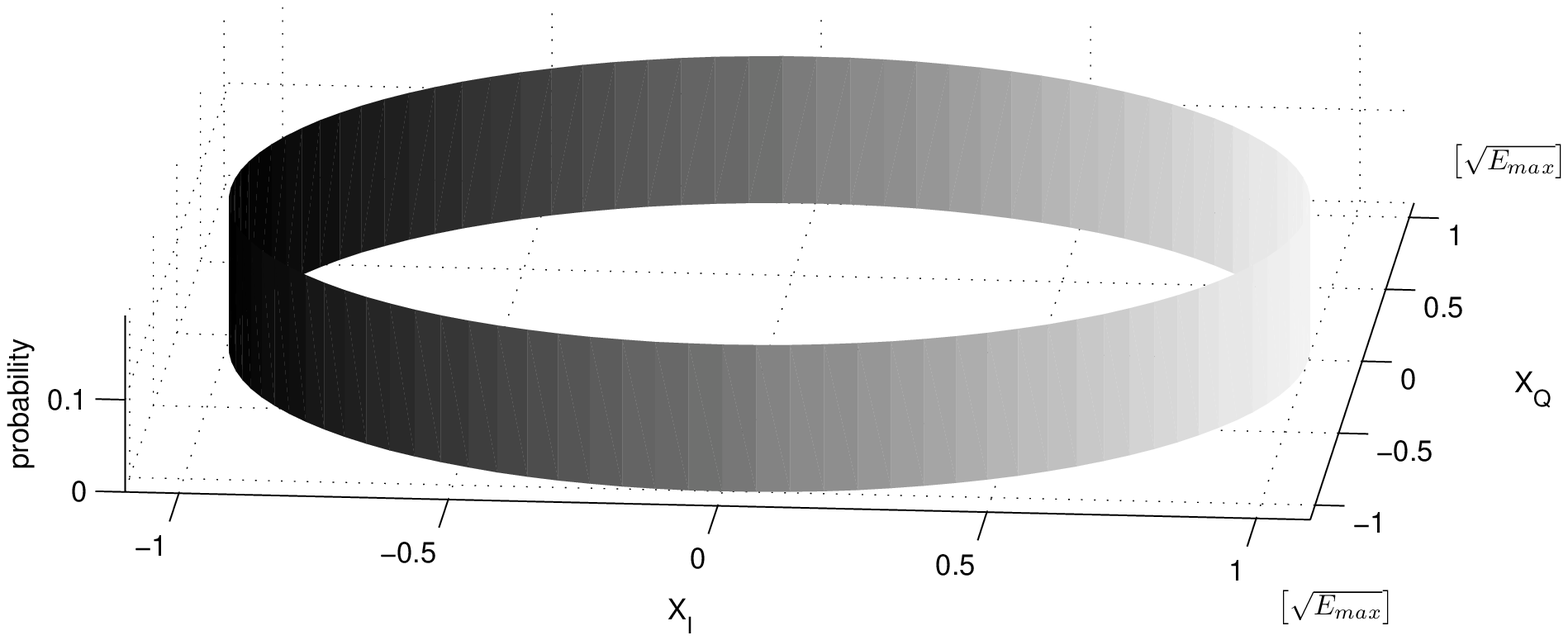}
    \label{fig:circle-CAD-b}
  }
  \subfloat[CAD when $\pSNR=16$.]{
    \includegraphics[width=0.45\figurewidth]{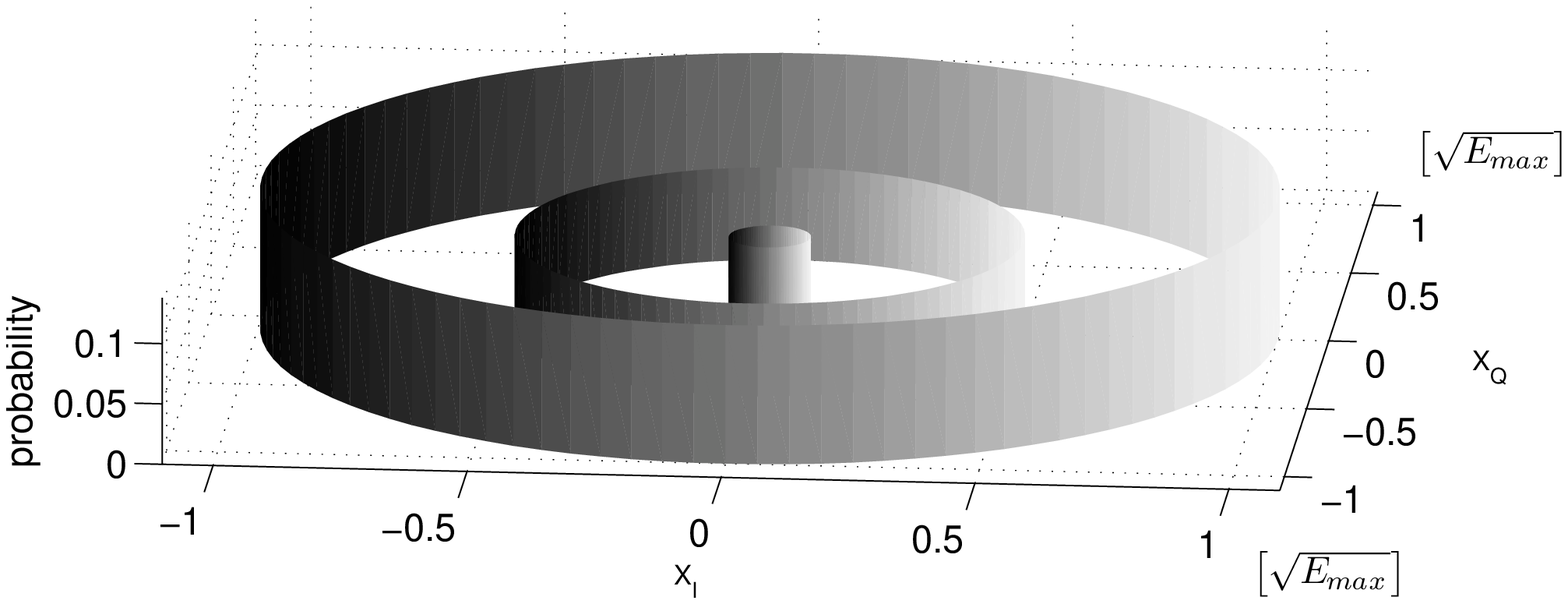}
    \label{fig:circle-CAD-c}
  }
  \caption{CADs of AWGN channel under circular constraint.}
  \label{fig:circle-CAD}
\end{figure}
Figure \ref{fig:circle-CAD} shows the numerically computed CADs under
the circular constraint $X_I^2+X_Q^2\le E_{\max}$. Figure
\ref{fig:circle-CAD-a} shows radial positions of probability masses of
the CADs. The number of the radial points is 1 for small enough peak
SNRs, while it increases as $\pSNR$ becomes larger. One of the points
is always located at the boundary $r=\sqrt{E_{\max}}$. Accordingly,
the CAD is a single circle for a small $\pSNR$
(Fig.~\ref{fig:circle-CAD-b}) and multiple concentric circles for a
larger $\pSNR$ (Fig.~\ref{fig:circle-CAD-c}).

\begin{figure}[htbp]
  \centering
  \subfloat[Constellation of 16APSK.]{
    \includegraphics[width=0.5\figurewidth]{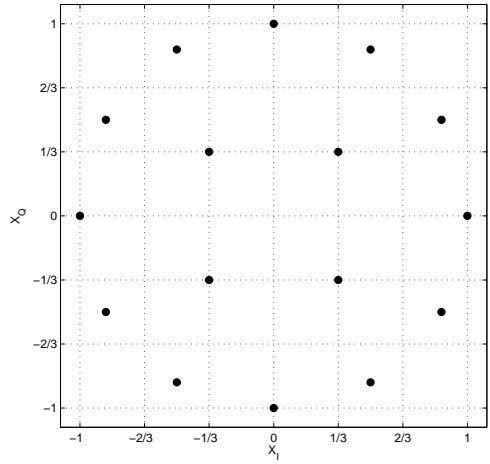}
    \label{fig:APSKconstellation}
  }\\
  \subfloat[Achievable rates with QPSK, 16PSK, and 16APSK compared
  with capacity under circular constraint.]{
    \includegraphics[width=0.8\figurewidth]{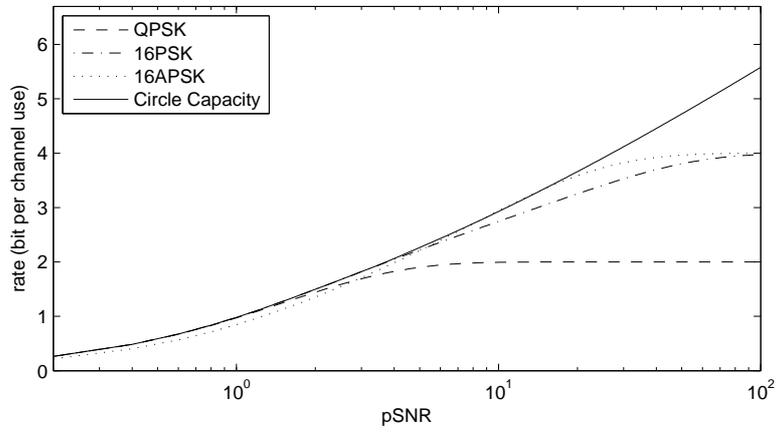}
    \label{fig:APSKcapacity}
  }
  \caption{Constellation and achievable rate with APSK.}
  \label{fig:constellation and capacity of APSK}
\end{figure}

The achievable rates with different constellations are shown along
with the capacity under the circular constraint in
Fig.~\ref{fig:APSKcapacity}. The achievable rate with QPSK is very
close to the capacity for small $\pSNR$s. One also observes that
16PSK, although not popular in current communications systems, has the
achievable rate closer to the capacity up to a moderate value of
$\pSNR$. Increasing the number $n$ of signal points in $n$PSK makes
the achievable rate closer to the capacity up to a yet larger $\pSNR$
value, but the rate becomes falling off from the capacity beyond that
$\pSNR$ value (Fig.~\ref{fig:APSKcapacity}). Figure
\ref{fig:circle-CAD-a} explains the reason. As the number $n$ of
$n$PSK increases, the input distribution approaches a single circle,
while the number of the circles increases for the CAD.

As is the case under the box constraint, one can expect that a higher
rate should be achievable by designing the input distribution so as to
make it similar to the CAD under the circular constraint. The shapes
of CADs under the circular constraint imply that amplitude and phase
shift keying (APSK)-type modulations work better than PSKs for a
larger $\pSNR$. As an example, we consider 16APSK whose constellation
is shown in Fig.~\ref{fig:APSKconstellation}, which is intended to
mimic the CAD with $\pSNR\sim10$ consisting of two circular
components.
The achievable rate with 16APSK is compared with those of PSKs and the
capacity under the circular constraint in
Fig.~\ref{fig:APSKcapacity}. As we have expected, the achievable rate
with 16APSK is worse than those of PSKs for a $\pSNR$ less than around
5 but is very close to the capacity under the circular constraint for
a larger $\pSNR$ up to around 16. Note that $\pSNR$ of 5 corresponds
to the point where the number of the circles of the CAD increases from
1 to 2 in Fig.~\ref{fig:circle-CAD-a}. Thus we expect that the APSK
with more amplitude shifts would have very good achievable rates for a
larger $\pSNR$ and that curves like those shown in
Fig.~\ref{fig:APSKcapacity} would indicate the corresponding $\pSNR$
to switch between them.

\section{Conclusion}

When designing constellation in digital communication, it is important
to properly take into consideration practical constraints such as peak
power constraint, but capacities of channels have conventionally been
argued in terms of average SNR, and peak power constraints have mostly
been considered indirectly via PAR. In this paper, we have proposed a
direct approach to compare achievable rates and capacity of a channel
under practical constraints on input. To demonstrate significance of
the proposed approach, we have studied, as an example, a discrete-time
complex-valued AWGN channel under peak power constraint, and have
shown how close achievable rates with commonly used PSKs and QAMs are
to the capacity under peak power constraint. In order to accomplish
it, the achievable rates with PSKs and QAMs for a complex-valued AWGN
channel have been evaluated and aligned according to the peak power,
and compared with the capacities numerically evaluated under the two
different forms (box and circular) of peak power constraint.

We have observed that the achievable rates with $n$QAM are very close
to the capacity under the box constraint for some range of $\pSNR$,
and that the range shifts to larger $\pSNR$ as the modulation level of
$n$QAM increases. We have also observed that the achievable rates with
$n$PSK are very close to the capacity under the circular constraint
for small $\pSNR$. Our results have also suggested that APSK-type
modulation is expected to have an achievable rate close to the
capacity under the circular constraint for larger $\pSNR$. The
achievable rate with 16APSK has been computed to support this
expectation. These results show that appropriate design of practical
discrete adaptive modulation will bring us a very efficient
modulation.

In this paper, we have considered only the case with a single (peak
power) constraint on input. As an extension, we can consider more
complicated forms of constraints, such as the cases in which both
average power and peak power are simultaneously
constrained~\cite{Smith1971informcontrol,ShamaiBarDavid1995ieeeit}.
It is straightforward to study such complicated cases by rewriting the
conditions, and similar results will be observed. This is because the
CAD becomes discrete in many cases for many types of channels and
constraints. Similar comparisons between achievable rates and
capacities under those conditions will provide useful guidelines for
adaptive modulations. This could not be possible with the indirect
approach in which the capacity is derived in terms of average power of
input and constraints are discussed separately.

We have also restricted our discussion to the case with discrete-time
input. To the best of our knowledge, there has been no direct
comparison in the literature between the achievable rates with
different constellations and the capacity under peak power constraint
even in the case with a discrete-time AWGN channel, and as we have
demonstrated, the direct comparison for the discrete-time channel has
yielded several novel quantitative observations regarding gaps between
achievable rates with practical constellations and the capacity under
peak power constraint. On the other hand, it has now been a common
practice in the indirect approach to study the PAR in continuous time
domain, typically in terms of the complementary cumulative
distribution function (CCDF) of PAR values. Another important
direction of extending our analysis is therefore to consider peak
power constraint on continuous-time input in the direct approach as
well.

Finally, we would like to mention that it has been shown from results
in Bayesian statistics~\cite{BergerBernardo1992bayesian} that the CAD
for an AWGN channel under the box constraint should approach the
Jeffreys prior as $\pSNR$ becomes
infinity~\cite{ClarkeBarron1994jspi,Zhang1994phd}, which in this case
is a uniform distribution. This observation directly implies that
$n$QAM and well designed $n$APSK with a large $n$ would be a good
approximation of the CAD under the peak power constraint for large
$\pSNR$s.

\end{document}